\magnification 1200
\font\tenmsb=msbm10   
\font\sevenmsb=msbm7
\font\fivemsb=msbm5
\newfam\msbfam
\textfont\msbfam=\tenmsb
\scriptfont\msbfam=\sevenmsb
\scriptscriptfont\msbfam=\fivemsb
\def\Bbb#1{\fam\msbfam\relax#1}
\let\nd\noindent 
\def\qed{\hbox{\hskip 6pt\vrule width6pt height7pt depth1pt \hskip1pt}}
\def\natural{{\rm I\kern-.18em N}}
\def\A{{\cal A}}
\def\B{{\cal B}}
\def\W{{\cal W}}

\def\integer{{\rm Z\kern-.32em Z}}
\def\chix{{\raise.5ex\hbox{$\chi$}}}
\def\Z{{\Bbb Z}}
\def\real{{\rm I\kern-.2em R}}
\def\R{{\Bbb R}}
\def\complex{\kern.1em{\raise.47ex\hbox{
            $\scriptscriptstyle |$}}\kern-.40em{\rm C}}

\def\vs#1 {\vskip#1truein}
\def\hs#1 {\hskip#1truein}

\def\Month{\ifcase\number\month \relax\or January \or February \or
  March \or April \or May \or June \or July \or August \or September
  \or October \or November \or December \else \relax\fi }
\def\date{\Month \the\day, \the\year}


  \hsize=6truein        \hoffset=.25truein 
  \vsize=8.8truein      
  \pageno=1     \baselineskip=13.9pt
  \parskip=4 pt         \parindent=20pt
  \overfullrule=0pt     \lineskip=0pt   \lineskiplimit=0pt
  \hbadness=10000 \vbadness=10000 
\pageno=0

\footline{\ifnum\pageno=0\hss\else\hss\tenrm\folio\hss\fi}
\hbox{}
\vskip 1truein\centerline{{\bf ISOMORPHISM OF HIERARCHICAL STRUCTURES}}
\vskip .3truein\centerline{by}
\vskip .2truein
\centerline{{Charles Radin${}$
\footnote{*}{Research supported in part by NSF Grant No. DMS-9531584 and
\hfill\break \indent Texas ARP Grant 003658-152\hfil}\ \ 
and\ \ Lorenzo Sadun${}$
\footnote{**}{Research supported in part by NSF Grant No. DMS-9626698 and
\hfill\break \indent Texas ARP Grant 003658-152\hfil}}}
\vskip .2truein\centerline{\vbox{
Department of Mathematics, University of Texas, Austin, TX\ \ 78712}}
\vs.5
\centerline{{\bf Abstract}}
\vs.1 \nd
We consider hierarchical structures such as Fibonacci sequences and
Penrose tilings, and examine the consequences of different choices for
the definition of isomorphism. In particular we discuss the role such
a choice plays with regard to matching rules for such structures.
\vs1
\vfill\eject 
\nd
{\bf 0. Introduction}
\vs.1
This paper concerns certain hierarchical tilings of Euclidean space,
of which a well known example is a ``kite \& dart'' tiling of the
plane [Gar]. Such tilings are noted for two properties: their unusual
rotational symmetries (``ten-fold symmetry''), and the fact that their
global structures derive from local (``matching'') rules.  We will
examine the consequences of different choices for the definition of
isomorphism between such structures. To be more specific, we assume
that any tiling of interest is embedded in a compact metric space of
tilings on which the connected component $G$ of the Euclidean group
(or sometimes just its translation subgroup) acts in the natural way, 
and has a unique translation invariant Borel probability measure.
We consider two specific notions of isomorphism: we say the tiling
spaces $X$ and $Y$ are ``topologically conjugate'' if there is a
homeomorphism between the spaces intertwining the actions of the group; and
we say $X$ and $Y$ are ``measurably conjugate'' if there are subsets
$Z_X\subset X$ and $Z_Y\subset Y$, of measure zero with respect to 
the unique translation invariant measure and invariant under
the action of the group, and a bimeasurable map from $X/Z_X$ onto $Y/Z_Y$
intertwining the actions of the group. We investigate the consequences of
assuming one or the other of these notions of isomorphism for such
tilings.

In [RaS] we analyzed the rotational symmetry of such hierarchical
structures, and found a way to associate a useful invariant for
measurable conjugacy; the methods used would not have allowed for
topological conjugacy, though perhaps other methods might. In this
paper we analyze hierarchical structures from the point of view of their
other main characteristic, the fact that their global structures
derive from local rules, and here we show that topological conjugacy
is definitely not as broadly applicable as measurable conjugacy.

The above approach to analyzing tilings is a natural outgrowth of the
way symbolic sequences or arrays are studied, through embeddings in
symbolic dynamical systems. The specific matter we consider for
tilings, concerning local rules, is actually better understood in that
older context. So we begin with more detailed descriptions of the
relevant ideas within symbolic arrays.

One result of this paper is the sharpening of a remarkable result of
Shahar Mozes [Moz].  Mozes showed that every substitution subshift
with $\Z^2$ action, satisfying mild hypotheses, is measurably conjugate 
via an explicit
construction to a uniquely ergodic subshift of finite type. 
(Such a result is commonly
said to provide ``matching rules'' for the functions in the
substitution subshift.) The map Mozes constructs is 
continuous (although not uniformly continuous) after exclusion of appropriate 
sets of measure zero. 
A natural question is whether such exclusion is an artifact
of the proof or an essential feature of the theorem: Can Mozes'
theorem be strengthened to give a topological conjugacy?  We answer
this question in the negative, exhibiting explicit substitution
subshifts with $\Z^2$ action which are not topologically conjugate to
any subshift of finite type.

After considering the simpler symbolic situation we explore the
analogous questions for tiling dynamical systems with $\R^d$ actions.
Mozes' proof can probably be extended without major difficulty to
subshifts in higher dimensions and to tiling systems in which each
polyhedron only appears, in any single tiling, in only finitely many
orientations. However, more general tiling systems, such as the
pinwheel in $\R^2$ [Ra1] and quaquaversal tilings in $\R^3$ [CoR],
require a significant extension of the proof in [Moz]. This was done
for the pinwheel in [Ra1] and then in general in [GoS].  In each case
a measurable conjugacy is constructed, but the map is only continuous
after exclusion of certain sets of measure zero.  We demonstrate, as with subshifts,
that these theorems cannot be strengthened to the topological
category: there exist substitution tiling systems that are measurably
conjugate to finite type tiling systems but which are not
topologically conjugate to any finite type tiling system.

In both settings, with subshifts and tilings, we give conditions,
easily satisfied in natural examples, for which such a topological
conjugacy is not possible.  However, the two constructions are somewhat
different for there are important properties of subshifts whose
natural analogues do not hold for tilings. For instance, for subshifts
it is well known that every topological conjugacy is a sliding block
code while for tilings we show this to be false.

This last point is both surprising and profound.  The continuous
nature of the $\R^d$ actions allows for two kinds of small changes to
a tiling.  Tiles may be changed near infinity, or a small Euclidean
motion may be applied to the entire tiling.  Neither notion is
topologically invariant since, as we show, a topological conjugacy can
encode pattern information from distant regions as small Euclidean
motions near the origin.  This mechanism, which has no analogue in the
theory of subshifts (where the translation group acts discontinuously),
indicates that the nature of topological conjugacies of tilings is
much subtler than topological conjugacy of subshifts.

In section 1 we discuss subshifts and show that many substitution
subshifts are not topologically conjugate to finite type subshifts.
In section 2 we explore topological conjugacy in tiling systems,
giving an example of a conjugacy that is not a sliding block code.
This example also demonstrates that local finiteness, a property
automatic for subshifts, is not a topological invariant for
tilings. In section 3 we return to the question of Mozes' theorem,
only now for tilings, and show there exist substitution tilings that
are measurably conjugate, but not topologically conjugate, to finite
type tilings.

\vs.1
\nd {\bf 1. Notation and results for subshifts}
\vs.1
For $n\ge 0$ and finite abstract alphabet $\A$ (always assumed
nonempty) let $\Pi_n$ be the restriction of functions in $\A^{\Z^d}$
onto $[-n,\ n]^d$. Let $\sigma_\A^j$ denote the shift by $j\in \Z^d$
on (any subshift of) $\A^{\Z^d}$: $(\sigma_\A^j x)_k=x_{k+j}$.
For each $n\ge 1$ define the set 

$$X_n=\{x\in
\A^{\Z^d}:\Pi_n [\sigma_\A^jx]\in \Pi_n[X]\hbox{ for all }j\in \Z^d\}.
\eqno 1)$$ 
That is, $X_n$ is the set of functions that, restricted to cubes of
size $n$, look like elements of $X$. 
We say the subshift $X$ is ``of finite type'' if $X=X_n$ for some
$n\ge 1$.

Given any second alphabet $\B$ and a ``block map''
$\Phi:\Pi_n(X)\longrightarrow \B$ we can define the ``sliding block
code (of code size $n$)'' $\phi:X\to \B^{\Z^d}$ by
$\phi(x)_j=\Phi(\Pi_n[\sigma^j_\A(x)])$. Note that $\phi$ is
continuous and intertwines the shifts on $X$ and $\phi(X)\subset
\B^{\Z^d}$; if it is also invertible it is a topological conjugacy.
An equivalent description of a sliding block code of size $n$ is a
map $X \to Y$, intertwining the $\Z^d$ actions, such that, 
whenever $\Pi_n x = \Pi_n x'$, then $\Pi_0 \phi(x) = \Pi_0 \phi(x')$.

\nd
{\bf Theorem 1}. (Curtis-Lyndon-Hedlund; [LiM]). Any topological
conjugacy $\phi$ between $\Z^d$ subshifts $X$ and $Y$ is a sliding block code.

\nd Proof: Let $d_X$ and $d_Y$ be the metrics on $X$ and $Y$, respectively.
Since $X$ and $Y$ are compact metric spaces, any continuous
map between them is uniformly continuous. We can therefore pick constants
$\epsilon$, $\delta$ and $n$ such that $\Pi_n x = \Pi_n x'$ implies
$d_X(x,x')<\delta$, which implies $d_Y(\phi(x),\phi(x'))<\epsilon$,
which implies $\Pi_0 \phi(x)=\Pi_0 \phi(x')$. \qed

The following theorem seems to be widely known but does not appear to
be in the literature. 
\vs.1 \nd 
{\bf Theorem 2}. Assume the $\Z^d$ subshift $X$ is of finite type and 
topologically conjugate to the subshift $Y$. Then $Y$ is of finite
type.
\vs.1
\nd Proof. Assume $X=X_n$ and $\phi:Y\longrightarrow X$ is the 
topological conjugacy.  By Theorem 1 both $\phi$ and $\phi^{-1}$ are
sliding block codes, say of size $m$ and $m'$, so there exists a block
map $\Phi:\Pi_m[Y]\to \A$ such that
$\phi(y)_j=\Phi(\Pi_m[\sigma_\B^j(y)])$. For any $p\ge m$ we can
extend $\phi$ to $ \phi':Y_p\to \A^{\Z^d}$ by this rule. Notice that the
image of $\phi'$ is in $X_{p-m}$, so, for $p \ge m+n$, $\phi'$ maps
$Y_p$ to $X$.

Now fix $p = \max( m+m', m+n)$ and let $\rho = \phi^{-1} \circ \phi': Y_p
\longrightarrow Y$.  Since $\rho$ is the product of sliding block codes
of size $m$ and $m'$, it is itself a sliding block code of size
$k=m+m'$.  $\rho$ is the identity when
restricted to $Y$; since $k \le p$, $\Pi_k Y_p = \Pi_k Y$, 
so the map $\rho$ is the identity on $Y_p$, implying  $Y_p=Y$. \qed

We next define substitution subshifts and show that certain substitution
subshifts are not of finite type.  By Theorem 2 they cannot be 
topologically conjugate to subshifts of finite type, but by Mozes' theorem
they are measurably conjugate to subshifts of finite type.

We begin in dimension 1. Given an alphabet $\A$ we define the set of
``words'' as $\W=\cup_{n\ge 1}\A^n$. We assume given a ``substitution
function'' $\psi :\A \to \W$ such that $\psi(a)\in \cup_{n\ge 2}\A^n$
for some $a\in \A$. A word of the form $\psi^n(a)$ is said to be a
``letter of level $n$'' for any $n\ge 0$, and we denote by $\W^\psi$
the set of all such words.  Finally, we define the substitution
subshift associated with the substitution $\psi$ as
$$
\{ 
x\in \A^\Z |\hbox{ For all }j\in\Z,\ k\ge 0\ \hbox{ there exists }
w \in \W^\psi \hbox{ s.t. }
(x_j,x_{j+1},\cdots , x_{j+k})\subseteq w 
\} . \eqno 2)
$$ 
It is a simple fact that every function in such a subshift can
simultaneously be considered a function with values in $\psi^n(\A)$
for each value of $n\ge 0$. In interesting cases functions in a
substitution subshift can be considered functions with values in
letters of each higher level in only one way; such subshifts are said
to be ``uniquely derivable''. (This is the origin of the term
``hierarchical'' for these structures. The same hierarchical
phenomenon exists for tilings, and in a manner easier to understand,
which we illustrate in Figure 2.)

The above has a straightforward generalization to $\Z^d$ subshifts as
long as the images of the substitution function fit together
geometrically, so that one can iterate the substitution. This is
automatic for $\Z^d$ substitution subshifts that are products
of $\Z$ substitution subshifts, which we illustrate by an example
below.

\nd {\bf Corollary 1}. If a $\Z^2$ substitution subshift $X$ contains a
function $x$ such that $x_{j,k}=x_{j,k+1}=x_{j+1,k}=x_{j+1,k+1}$
for some $(j,k)\in \Z^2$, and $X$ is topologically conjugate to a finite
type subshift, then $X$ contains a periodic function.
\vs.1 \nd
Proof. The condition on $x$ defines a word $C\in \A^{[0,1]^2}$.
Consider the large words $W\in \A^{[0,N-1]\times[0,M-1]}$ produced
when the substitution is applied repeatedly to $C$, and consider the
periodic function $p$ defined by the condition $(\sigma^{Nj,Mk}_\A p)_{[0,N-1]\times
[0,M-1]}=W$ for all $j,k\in\Z$. Since $X$ is topologically conjugate
to a subshift of finite type, by Theorem 2 there is some $n\ge
1$ such that $X=X_n$. Taking $W$ large enough it follows that $X$
contains the periodic function $p$. \qed

\nd {\bf Corollary 2}. There exists a $\Z^2$ substitution subshift
that is measurably conjugate to a finite type subshift, but is not 
topologically conjugate to any finite type subshift.

\nd Proof.
Consider the uniquely derivable symbolic Fibonacci 
substitution defined,
using the alphabet $\A=\{a,b\}$, by the substitution function
$\psi(a)=b,\ \psi(b)=a b$; let $F(1)$ be the strictly ergodic
substitution subshift thus determined. Let $F(2)\subset\{\A\times
\A\}^{\Z^2}$ be the substitution subshift which is the product of
$F(1)$ with itself:
$$ 
\eqalign{ \psi_2(a\times a)=(b \times b); & \qquad 
\psi_2(b\times a)=\pmatrix{a\times b & b\times b}; \cr
\psi_2(a\times b)=\pmatrix{b\times b\cr b\times a}; & \qquad
\psi_2(b\times b)=
\pmatrix{a\times b & b\times b \cr a\times a & b\times a}.} \eqno 3)
$$
It follows [Moz] that $F(2)$ is uniquely derivable. 
Mozes' theorem states 
that under rather general conditions (satisfied by
$F(2)$) a $\Z^2$ substitution subshift $X$ which is uniquely derivable
is measurably conjugate to some finite type subshift, so $F(2)$ is
measurably conjugate to a finite type subshift.

A simple calculation, however, shows that $\psi_2^3(b\times b)$, and
hence all functions in $F(2)$, contain
the block $\pmatrix{ b\times b & b\times b \cr b\times b & b\times
b}$, and a standard argument ([Ra2], [Ra3]) using the unique
derivability of $F(2)$ implies that $F(2)$ contains no periodic
functions. Corollary 1 then implies that $F(2)$ is not topologically
conjugate to any finite type subshift. \qed
\vs.1
\nd {\bf 2. Conjugacies of Fibonacci Tilings}
\vs.1
Now we consider ``tiling systems''. Given a finite collection $\A$ of
polyhedra in $d$ dimensions, we define $X^\A$ as the space of all
tilings of Euclidean $d$-space by congruent copies of elements of
$\A$, that is, congruent under the connected Euclidean group. 
(We assume $X^\A$ is nonempty.) We put a metric $d(\cdot,\cdot)$
on $X^\A$ as follows.
$$
d(x, y)\equiv \sup_{n}{1\over n}m_{H}[B_{n}(\partial x),B_{n}(\partial
y)], \eqno 4)
$$
where ${B}_{n}(\partial x)$ denotes the intersection of two sets:
the closed ball ${B}_n$ of radius $n$ centered at the origin of the
Euclidean space and the union $\partial x$ of the boundaries $\partial
a$ of all tiles $a$ in $x$. $m_{H}$ is the Hausdorff metric on compact
sets defined as follows. Given two compact subsets $A$ and $B$ of $\R^m$,
$m_{H}[A,B] = \max
\{ {\tilde d} (A,B), {\tilde d} (B,A)\}$, where
$${\tilde d} (A,B) =  \sup_{a \in A}\inf_{b\in B} ||a - b||, \eqno 5)
$$
with $||w||$ denoting the usual Euclidean norm
of $w$. (We assume the tiles are small enough so that ${B}_{1}(\partial x)$
is nonempty for any $x$.)
It is not hard to show that with this metric $X^\A$ is compact
and that the natural representation of the connected component of
the Euclidean group on $X^\A$ is continuous. Finally,
let $\hat G$ be either the group of translations of $\R^d$ or the
connected component of the Euclidean group.  A tiling system is a
closed $\hat G$-invariant subset of $X^\A$. 

Two important properties of tiling systems are ``local finiteness''
and ``finite type''.  For each $R>0$ and polyhedron $a\in \A$
appearing in the tiling $x$ of $X$ consider the set $H(R,x,a)$ of
polyhedra in $x$ which intersect the open ball of radius $R$ centered
at the center of mass of $a$. Define $H(R,X)$ to be the union of all
such ``neighborhoods of radius $R$'' for all $a \in x$ and $x \in
X$. A tiling system $X$ is ``locally finite'' if, for every $R>0$, the
set $H(R,X)$ is finite up to the action of $\hat G$.  In practice, tiling
systems are almost always assumed to be locally finite.

Now, in analogy to the subshifts $X_n$, we define $X_R$ to be the set
of all $x$ in $X^\A$ such that $x$ only has neighborhoods of radius
$R$ which are in $H(R,X)$.  A tiling system $X$ is said to be ``of
finite type'' if $X=X_R$ for some $R>0$.

We will be considering the analogue among tiling systems of the
substitution subshifts. Traditionally such substitution tiling systems
are defined through a substitution function as for subshifts, whereby
for each letter (polyhedron) there is given a way to decompose it into
polyhedra similar to those in the alphabet, all smaller by some factor
$\gamma<1$, so that following the decomposition by a stretch about the
origin by $1/\gamma$ one associates to each letter of $\A$ a
letter of level 1, and so on. For example, Figure 1 shows a
substitution for a ``chair'' tile, in this case with $\gamma=1/2$, and
Figure 2 shows the hierarchical structure in a piece of a chair
tiling.  Applying the substitution to all tiles in a tiling gives an
automorphism $\psi: X \to X$. Notice that $\psi$ commutes with
rotations about the origin and that $\psi \circ
\sigma^\alpha = \sigma^{\alpha/\gamma}\circ \psi$, where $\sigma^\alpha$ 
is a translation by $\alpha \in \R^d$.

We use this last property to define a generalized notion of
substitution tiling system. For our purposes, a substitution tiling
system is a tiling system with an automorphism $\psi$ and a constant
$\gamma < 1$ such that $\psi$ commutes with any rotations about the
origin in $\hat G$
and such that $\psi \circ
\sigma^\alpha = \sigma^{\alpha/\gamma}\circ \psi$ for all $\alpha \in \R^d$.

A simple example of a substitution tiling is the standard Fibonacci
tiling in one dimension.  In this tiling the alphabet consists of two
tiles (intervals), $A$ and $B$, of length $|A|=1$ and
$|B|=\tau=[1+\sqrt{5}]/2$.  The substitution consists of expanding
about the origin by a factor of $\tau$, and then replacing each
expanded $A$ (of size $\tau$) with a $B$, and replacing each expanded
$B$ (of size $\tau^2=1+\tau$) with an $A$ and a $B$.  The resulting
space of tilings is precisely the set of tilings by $A$'s and $B$'s
arrayed in the sequence of the Fibonacci subshift $F(1)$.

This last characterization allows us to define general Fibonacci
tilings, based on any two tiles $A$ and $B$, as the set of tilings in
which the pattern of $A$'s and $B$'s is the same as a sequence in the
Fibonacci subshift.  If $|B| \ne \tau |A|$ this is not a
substitution tiling in the traditional sense, as the substitution
$A \to B, B \to AB$ is not an expansion by a constant factor.  However,
it is a substitution tiling system in the extended sense of this paper. 

\nd {\bf Theorem 3}.  Let $X$ and $Y$ be Fibonacci tiling systems, 
with tiles $A_X, B_X$ and $A_Y, B_Y$, respectively.  If $|A_X|+\tau |B_X|=
|A_Y| + \tau |B_Y|$ then $X$ and $Y$ are topologically conjugate.
Moreover, the conjugacy is {\it not} a sliding block code.

\nd Proof: We define a map $\phi: X \longrightarrow Y$ as follows.
A tiling $x \in X$ defines an element of $F(1)$ (up to translation),
which in turn defines a tiling $y \in Y$, up to translation.  To fix
the translation we pick a tile $a \in x$. This tile $a$ is a letter of
level 0, which sits inside a letter $a_1$ of level 1, which sits inside
a letter $a_2$ of level 2, and so on. We approximate the placement of
the tiles in $y$ by lining up the middle of the letter in $y$ that
corresponds to $a_n$ with the middle of $a_n$.
Since $|A_X|+\tau |B_X|= |A_Y| + \tau
|B_Y|$, the letters of level $n$ in the two tilings differ in size by
a constant times $\tau^{-n}$.  The adjustment in translation between
the $n$th approximant and the $n+1$st is then exponentially small, 
and we can take the limit as $n \to
\infty$ to obtain $\phi(x)$. Loosely speaking, the map $\phi$ lines up
the corresponding letters of infinite level.

We must show that this map is well defined, in that it does not depend
on the choice of tile $a$.  If $a' \in x$ is another tile, then there are
two possibilities.  The more common possibility is that there is a level $n$ at which
$a_n=a'_n$,  so the approximants to $\phi(x)$ based on $a'$ are eventually
the same as the approximants based on $a$.  Alternatively, there may be 
a point between $a$ and $a'$ which is the endpoint of letters of arbitrarily
high level.  It is not hard to check that, working from either $a$ or $a'$,
$\phi$ aligns this special point in $x$ with the corresponding special
point in $\phi(x)$. 

The map $\phi$ is clearly a bijection which commutes with
translations.  It is continuous since a small change to a tiling $x$, either
via a small translation or by changing the tiles near infinity, results
is a small change to $\phi(x)$.  A small translation in $x$ turns into
a small translation in $\phi(x)$, while changing the distant tiles of $x$
changes the distant tiles of $\phi(x)$, and also causes a small translation.

This last point means that $\phi$ is not a sliding block code.  If two
tilings $x$ and $x'$ agree exactly on a large neighborhood of the
origin but disagree near infinity, then the sequence of tiles of
$\phi(x)$ and $\phi(x')$ will agree near the origin, but the placement
of the tiles will differ by a small translation.  Thus there is no 
neighborhood of the origin where $\phi(x)$ and $\phi(x')$ agree exactly.
\qed

Let $|A| + \tau |B|$ have some fixed value.
There are two special Fibonacci tilings.  One is the standard Fibonacci
tiling with $|B|=\tau|A|$.  The other has $|B|=|A|$, and thus closely
resembles the Fibonacci subshift $F(1)$.  By Theorem 3 any Fibonacci
tiling is conjugate to each of these.  In particular,

\nd {\bf Corollary 3}. Every Fibonacci tiling system is a substitution
tiling system.

\nd Proof: Every Fibonacci tiling system is topologically conjugate
to a Fibonacci system $X$ with $|B_X|=\tau |A_X|$. As noted earlier,
$X$ is a substitution system with automorphism $\psi_X$ and scale
factor $\gamma=1/\tau$.  If $Y$ is another
Fibonacci tiling system and $\phi: X \longrightarrow Y$ is the conjugacy
of Theorem 3, then $\psi_Y = \phi \circ \psi_X \circ \phi^{-1}$ is
a substitution automorphism on $Y$. \qed

Unlike a sliding block code the domain of a topological conjugacy 
$\phi: X \longrightarrow Y$ of tiling systems
cannot always be extended to $X_R$ for $R$ sufficiently large.
Suppose for example that $X$ and $Y$ are Fibonacci tiling systems 
with $|A_X|=1$, $|B_X|=\tau$, $|A_Y|=2$ and $|B_Y|=1$. For any $R$ the
system $X_R$ contains periodic tilings, whose period $T$ is an integral linear
combination of 1 and $\tau$.  Since the tiles of $\A_Y=\{A_Y, B_Y\}$ have
integer length, there are no periodic tilings in $X^{\A_Y}$ with period $T$,
and so there are no maps $X_R \longrightarrow X^{\A_Y}$ that intertwine
translations.  

Since topological conjugacies of tilings are more general than sliding
block codes, topological invariants of subshifts are not automatically
topological invariants of tilings.  For example,

\nd {\bf Theorem 4}. There exist topologically conjugate tiling 
systems $X$ and $Y$ such that $X$ is locally finite and $Y$ is not.

\nd Proof: We work in 2 dimensions, with $\hat G$ being the 
translation group.  Given two square tiles $A_X$ and $B_X$ of unit
size, with edges parallel to the coordinate axes, let $X$ be the set
of all tilings such that the tiles meet full edge to full edge, and
such that each (horizontal) row of the tiling is a (1 dimensional)
Fibonacci tiling.  Let $\A_Y=\{A_Y, B_Y\}$, where $A_Y$ is a rectangle
of height 1 and width $\tau$ and $B_Y$ is a rectangle of height 1 and
width $\tau-1$.  Let $\phi$ be the conjugacy of Theorem 3, taking the
1 dimensional Fibonacci tiling with tile sizes 1 and 1 to the 1
dimensional Fibonacci tiling with sizes $\tau$ and $\tau-1$.  This map
is naturally extended to a map from rows of the tiling $X$ to
Fibonacci-like rows made up of the tiles of $\A_Y$.  Applying this to
every row of a tiling $x \in X$ defines an extended map (also called
$\phi$) from $X$ into $X^{\A_Y}$.  Let $Y$ be the image of this map.
Near the origin (or any other point), the rows of a tiling $y \in Y$
will appear shifted relative to one another.  Since there are an
infinite number of ways in which this shift can occur, $Y$ is not
locally finite. \qed
\vs.1
\nd {\bf 3. Conjugacies of tiling systems of finite type}
\vs.1
In this section tiling systems are assumed to be locally finite
unless stated otherwise.

A natural question, in the light of Mozes' theorem, is whether Theorem
2 can be extended to the category of tiling systems.  Certainly the
proof breaks down, as conjugacies of tilings are not necessarily
sliding block codes and need not extend to spaces $X_R$. However,
there is hope that the statement of the theorem remains true.

In this section we prove a somewhat weaker result, the analogue of
Corollary 1, thereby showing that there exist tiling systems that
are measurably conjugate to finite type systems but not topologically
conjugate. The key is showing that topological conjugacies, while not
sliding block codes, do preserve some useful geometrical information.
One can model a topological conjugacy $\phi:X\to Y$ between
tiling systems $X$ and $Y$ -- with metrics $d_X(\cdot,\cdot)$ and
$d_Y(\cdot,\cdot)$ and alphabets $\A$ and $\B$ -- by something close to a
sliding block code, as follows.

First we note that $\phi$ is uniformly continuous. So given
$\epsilon>0$ there is some $\delta >0$ such that
$d_Y[\phi(x),\phi(x')]<\epsilon$ provided $d_X[x,x']<\delta$. Now it
is known [RaW] that given any $\chix >0$ there is an $\epsilon >0$
such that if $d_Y[y,y']<\epsilon$ it follows that there is a rigid
motion $t$ within distance $\chix$ of the identity in $\hat G$ such
that the tilings $y$ and $\sigma^t_\B(y')$ ``agree to distance 1
around the origin'', in the sense that they match perfectly in their
polyhedra which intersect the ball of radius 1 centered at the origin.
Furthermore there is some $R>0$ such that if tilings $x$ and $x'$ in
$X$ agree to distance $R$ around the origin then $d_X[x,x']<\delta$
and so the tilings $\phi(x)$ and $\sigma^t_\B[\phi(x')]$ agree to
distance 1 around the origin for some $t\in \hat G$ of distance less
than $\chix$ from the identity. We have proven the following theorem.

\nd
{\bf Theorem 5}. Let $\phi$ be a topological conjugacy of the locally
finite tiling system $X$ in $\R^d$ onto the locally finite $\R^d$
tiling system $Y$. Given $\epsilon >0$ there is some
$R_\phi(\epsilon)>0$ such that if tilings $x$ and $x'$ in $X$ agree to
distance $R_\phi(\epsilon)$ around the origin then the tilings
$\phi(x)$ and $\sigma^t_\B[\phi(x')]$ agree to distance 1 around the
origin for some rigid motion $t$ of distance less than $\epsilon$ from
the identity in $\hat G$.

For any bounded connected subset $P$ of $\R^d$,
$R>0$, and tiling $x$, let $P^R$ be the union of open balls of radius
$R$ which intersect $P$. 

\nd
{\bf Corollary 4}. Let $\phi:X\to Y$ be a topological conjugacy of
locally finite tiling systems in $\R^d$. Then given $\epsilon >0$ and
a bounded connected subset $P$ of $\R^d$ there exists
$R_\phi(\epsilon) >0$ such that if tilings $x$ and $x'$ in $X$ agree
on $P^{R_\phi(\epsilon)}$ there is a rigid motion $t$ of distance less
than $\epsilon$ from the identity in $\hat G$ such that $\phi(x)$ and
$\sigma^t_\B[\phi(x')]$ agree on $P$.

\nd
Proof. The proof follows immediately by applying Theorem 5 to
all the balls of radius $R_\phi(\epsilon)$ which intersect $P$,
and noting that their overlapping regions lead to
consistency requirements in $Y$. \qed
 
We say that a tiling $x$ in an $\R^2$ tiling system $X$ with alphabet
$\A$ has a ``periodic
frame'' if there is some pair $\{s, t\}$ of linearly independent
translations for which $\sigma_\A^{as+bt}x$ and $\sigma_\A^{as+(b+1)t}x$
agree to distance $1$ about the origin for all $0\le a\le 1$ and
$|b|\le 1/8$, and also $\sigma_\A^{as+bt}x$ and $\sigma_\A^{(a+1)s+bt}x$
agree to distance $1$ about the origin for all $0\le b\le 1$ and
$|a|\le 1/8$.

\nd
{\bf Theorem 6}. If an $\R^2$ substitution tiling system $X$ contains
a tiling with a periodic frame, and $X$ is topologically conjugate to
a finite type tiling system, then $X$ contains a periodic tiling.

\nd
Proof. Assume $\epsilon >0$ is given and the frame is described by
with the defining notation above. By using the substitution repeatedly
if necessary we can assume without loss of generality that
$\min\{||ms+nt||,\ m,n\in \Z\}$ is larger than any given number. So we can
assume $x$ and $\sigma^t_\A x$ agree on $P_1^{R_\phi(\epsilon)}$ where
$P_1=\{as+bt:0\le a\le 1,\ |b|\le 1/8\}$ and that $x$ and $\sigma^s_\A
x$ agree on $P_2^{R_\phi(\epsilon)}$ where $P_2=\{as+bt:0\le b\le 1,\
|a|\le 1/8\}$. Applying Corollary 2 we see that $\phi(x)$ and
$\sigma^{t+\tau}_\B\phi(x)$ agree on $P_1$ and that $\phi(x)$ and
$\sigma^{s+\tau'}_\B\phi(x)$ agree on $P_2$ for rigid motions $\tau$
and $\tau'$ within distance $\epsilon$ of the identity of $\hat G$. Note
that because of the overlap between them, it follows that $\tau$ and
$\tau'$ are pure translations and so the central lines of $P_1$,
$\sigma_\B^{t+\tau}P_1$, $P_2$ and $\sigma_\B^{s+\tau'}P_2$ form a
parallelogram. And since $Y$ is of finite type this implies that it
contains a periodic tiling. But then so does $X$.\qed

Remark: At the end of section 1 we showed that the subshift with
$\Z^2$ action made as a product of the Fibonacci sequences with itself
could not be topologically conjugate to a finite type subshift.
Theorem 6 allows us to apply the same argument to tilings. Starting with any
Fibonacci tiling system we can define the product tiling system with
$\R^2$ action and show that it contains a tiling with a periodic
frame, and therefore cannot be topologically conjugate to a finite
type tiling system since it does not contain a periodic tiling. 
\vs.1
\nd
{\bf 4. Conclusion}
\vs.1
The best known tiling dynamical system is that of the Penrose kites
\& darts [Ra2, Ra3]. By this one refers to both a substitution system
and a finite type system, which are in fact topologically conjugate.
The fact that the conjugacy is topological has lead to interesting
work in noncommutative topology [AnP, Con].

To a large extent the interest in substitution tiling systems
is due to the fact that they are conjugate to finite type
systems. Theorem 6 shows that, when dealing with substitution tilings
and their conjugacy to finite type tilings, for reasonable generality
one cannot consider the tilings well defined up to topological
conjugacy -- though of course measurable conjugacy is sufficient. One
issue we have not resolved is whether or not the notion of finite type is 
itself a topological invariant among tiling systems.
\vs.1
\nd
{\bf Acknowledgements}
\vs.05 
\nd
We thank Klaus Schmidt for showing us the proof of
Theorem 2. After this work was distributed we learned of independent work
in progress [Pet], which has some overlap with ours concerning
Curtis-Lyndon-Hedlund type theorems for tiling dynamical systems.
\vs.2
{\centerline{\bf REFERENCES}}
\vs.2 \nd
[AnP]\ J.\ Anderson and I.\ Putnam, Topological invariants for substitution
tilings and their associated C$^\star$-algebras, {\it Ergodic\ Theory\ 
Dynam.\ Systems}\ 18 (1998), 509-537.
\vs.1 \nd 
[Con]\ A.\ Connes, {\it Noncommutative Geometry}, Academic Press, San Diego, 1994.
\vs.1 \nd 
[CoR]\ J.H.\ Conway and C.\ Radin, Quaquaversal tilings and rotations, 
{\it Invent. math.}\ 132 (1998), 179-188.
\vs.1 \nd 
[Gar]\ M.\ Gardner, Extraordinary nonperiodic tiling that enriches the
theory of tiles, {\it Sci.\ Am.\ (USA)} (December 1977), 110-119.
\vs.1 \nd 
[GoS]\ C.\ Goodman-Strauss, Matching rules and substitution tilings,
{\it Ann. of Math.}\ 147 (1998), 181-223.
\vs.1 \nd
[LiM]\ D.\ Lind and B.\ Marcus, {\it An Introduction to Symbolic
Dynamics and Coding}, Cambridge University Press, Cambridge, 1995.
\vs.1 \nd
[Moz]\ S.\ Mozes, Tilings, substitution systems and dynamical systems
generated by them, {\it J. d'Analyse Math.}\ 53 (1989), 139-186.
\vs.1 \nd
[Pet]\ K.\ Petersen, Factor maps between tiling dynamical systems, {\it Forum
Math.}\ 11 (1999), 503-512. 
\vs1 \nd
[Ra1]\ C.\ Radin, The pinwheel tilings of the plane,
{\it Ann. of Math.}\ 139 (1994), 661-702.
\vs.1 \nd
[Ra2]\ C.\ Radin, Miles of Tiles, Ergodic theory of $\Z^d$-actions, 
{\it London Math.\ Soc.\ Lecture Notes Ser.}\ 228
Cambridge University Press, Cambridge, 1996, pp.~237-258.
\vs.1 \nd
[Ra3]\ C.\ Radin, {\it Miles of Tiles}, Student Mathematical Library, Vol 1, Amer.\ 
Math.\ Soc., Providence, 1999.
\vs.1 \nd
[RaS]\  C.\ Radin and L.\ Sadun, An algebraic invariant of substitution tiling
systems, {\it Geometriae Dedicata}\ 73 (1998), 21-37.
\vs.1 \nd
[RaW]\ C.\ Radin and M.\ Wolff, Space tilings and local isomorphism,
{\it Geometriae Dedicata}\ 42 (1992), 355-360.
\vfill \eject

%
%
\newdimen\FigSize	\FigSize=.9\hsize 
%
\newskip\abovefigskip	\newskip\belowfigskip
\gdef\epsfig#1;#2;{\par\vskip\abovefigskip\penalty -500
   {\everypar={}\epsfxsize=#1\noindent
    \centerline{\epsfbox{#2}}}%
    \vskip\belowfigskip}%
%
\newskip\figtitleskip
\gdef\tepsfig#1;#2;#3{\par\vskip\abovefigskip\penalty -500
   {\everypar={}\epsfxsize=#1\noindent
    \vbox
      {\centerline{\epsfbox{#2}}\vskip\figtitleskip
       \centerline{\figtitlefont#3}}}%
    \vskip\belowfigskip}%
%
\newcount\FigNr	\global\FigNr=0
\gdef\nepsfig#1;#2;#3{\global\advance\FigNr by 1
   \tepsfig#1;#2;{Figure\space\the\FigNr.\space#3}}%
%
%
%
\gdef\ipsfig#1;#2;{
   \midinsert{\everypar={}\epsfxsize=#1\noindent
	      \centerline{\epsfbox{#2}}}%
   \endinsert}%
%
\gdef\tipsfig#1;#2;#3{\midinsert
   {\everypar={}\epsfxsize=#1\noindent
    \vbox{\centerline{\epsfbox{#2}}%
          \vskip\figtitleskip
          \centerline{\figtitlefont#3}}}\endinsert}%
%
\gdef\nipsfig#1;#2;#3{\global\advance\FigNr by1%
  \tipsfig#1;#2;{Figure\space\the\FigNr.\space#3}}%
\newread\epsffilein    
\newif\ifepsffileok    
\newif\ifepsfbbfound   
\newif\ifepsfverbose   
\newdimen\epsfxsize    
\newdimen\epsfysize    
\newdimen\epsftsize    
\newdimen\epsfrsize    
\newdimen\epsftmp      
\newdimen\pspoints     
\pspoints=1bp          
\epsfxsize=0pt         
\epsfysize=0pt         
\def\epsfbox#1{\global\def\epsfllx{72}\global\def\epsflly{72}%
   \global\def\epsfurx{540}\global\def\epsfury{720}%
   \def\lbracket{[}\def\testit{#1}\ifx\testit\lbracket
   \let\next=\epsfgetlitbb\else\let\next=\epsfnormal\fi\next{#1}}%
\def\epsfgetlitbb#1#2 #3 #4 #5]#6{\epsfgrab #2 #3 #4 #5 .\\%
   \epsfsetgraph{#6}}%
\def\epsfnormal#1{\epsfgetbb{#1}\epsfsetgraph{#1}}%
\def\epsfgetbb#1{%
%
%
\openin\epsffilein=#1
\ifeof\epsffilein\errmessage{I couldn't open #1, will ignore it}\else
%
%
   {\epsffileoktrue \chardef\other=12
    \def\do##1{\catcode`##1=\other}\dospecials \catcode`\ =10
    \loop
       \read\epsffilein to \epsffileline
       \ifeof\epsffilein\epsffileokfalse\else
%
%
          \expandafter\epsfaux\epsffileline:. \\%
       \fi
   \ifepsffileok\repeat
   \ifepsfbbfound\else
    \ifepsfverbose\message{No bounding box comment in #1; using defaults}\fi\fi
   }\closein\epsffilein\fi}%
%
%
\def\epsfsetgraph#1{%
   \epsfrsize=\epsfury\pspoints
   \advance\epsfrsize by-\epsflly\pspoints
   \epsftsize=\epsfurx\pspoints
   \advance\epsftsize by-\epsfllx\pspoints
%
%
   \epsfxsize\epsfsize\epsftsize\epsfrsize
   \ifnum\epsfxsize=0 \ifnum\epsfysize=0
      \epsfxsize=\epsftsize \epsfysize=\epsfrsize
%
%
     \else\epsftmp=\epsftsize \divide\epsftmp\epsfrsize
       \epsfxsize=\epsfysize \multiply\epsfxsize\epsftmp
       \multiply\epsftmp\epsfrsize \advance\epsftsize-\epsftmp
       \epsftmp=\epsfysize
       \loop \advance\epsftsize\epsftsize \divide\epsftmp 2
       \ifnum\epsftmp>0
          \ifnum\epsftsize<\epsfrsize\else
             \advance\epsftsize-\epsfrsize \advance\epsfxsize\epsftmp \fi
       \repeat
     \fi
   \else\epsftmp=\epsfrsize \divide\epsftmp\epsftsize
     \epsfysize=\epsfxsize \multiply\epsfysize\epsftmp   
     \multiply\epsftmp\epsftsize \advance\epsfrsize-\epsftmp
     \epsftmp=\epsfxsize
     \loop \advance\epsfrsize\epsfrsize \divide\epsftmp 2
     \ifnum\epsftmp>0
        \ifnum\epsfrsize<\epsftsize\else
           \advance\epsfrsize-\epsftsize \advance\epsfysize\epsftmp \fi
     \repeat     
   \fi
%
%
   \ifepsfverbose\message{#1: width=\the\epsfxsize, height=\the\epsfysize}\fi
   \epsftmp=10\epsfxsize \divide\epsftmp\pspoints
   \vbox to\epsfysize{\vfil\hbox to\epsfxsize{%
      \includegraphics{#1}%
      \hfil}}%
\epsfxsize=0pt\epsfysize=0pt}%
%
%
{\catcode`\%=12 \global\let\epsfpercent=
%
%
\long\def\epsfaux#1#2:#3\\{\ifx#1\epsfpercent
   \def\testit{#2}\ifx\testit\epsfbblit
      \epsfgrab #3 . . . \\%
      \epsffileokfalse
      \global\epsfbbfoundtrue
   \fi\else\ifx#1\par\else\epsffileokfalse\fi\fi}%
%
%
\def\epsfgrab #1 #2 #3 #4 #5\\{%
   \global\def\epsfllx{#1}\ifx\epsfllx\empty
      \epsfgrab #2 #3 #4 #5 .\\\else
   \global\def\epsflly{#2}%
   \global\def\epsfurx{#3}\global\def\epsfury{#4}\fi}%
%
%
\def\epsfsize#1#2{\epsfxsize}%
%
%

\epsfverbosetrue			
\abovefigskip=\baselineskip		
\belowfigskip=\baselineskip		
\global\let\figtitlefont\bf		
\global\figtitleskip=.5\baselineskip	
\nopagenumbers
\hbox{}
\vs2 
\vbox{\epsfig .7\hsize; 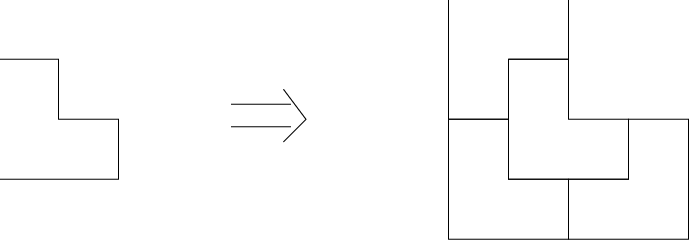;}
\vs1
\centerline{Fig.~1. The chair substitution}
\vfill\eject
\hbox{}
\vs.1 
\vbox{\epsfig .7\hsize; 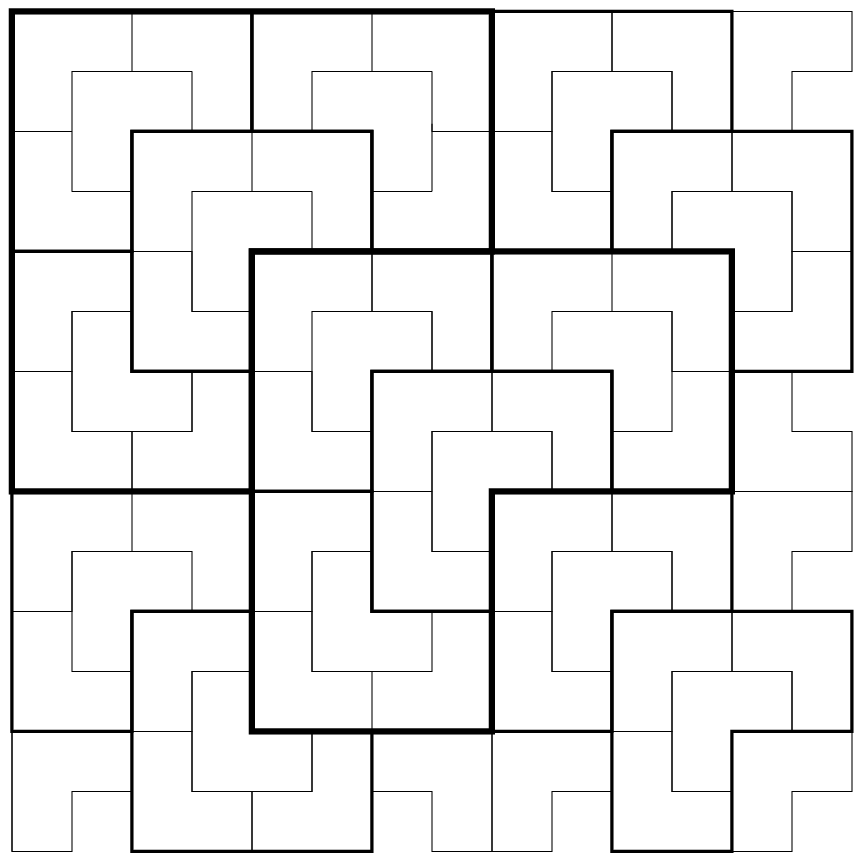;}
\vs1
\centerline{Fig.~2. 63 letters of level 0, 14 letters of level 1 and 2 letters of
level 2}
\centerline{in part of a chair tiling}
\vfill
\end

\end